\documentclass[a4paper,12pt]{article}
\usepackage{axodraw4j}
\usepackage{amsfonts,color}
\usepackage{amssymb,amsmath}
\usepackage{hyperref}
\usepackage{epsfig,graphicx}
\begin{document}
\thispagestyle{empty}

\begin{center}
\end{center}

\begin{center}

\vspace{1.7cm}

{\Large\bf{ 
Neutral Triple Vector Boson Production in Randall-Sundrum Model at the LHC}}

 \vspace{1.4cm}
 Goutam Das~\footnote{goutam.das@saha.ac.in}\ ,
 \hspace{.2cm} 
 Prakash Mathews~\footnote{prakash.mathews@saha.ac.in}
 \\[.5cm]

 Saha Institute of Nuclear Physics,\\
 1/AF Bidhan Nagar,\\
  Kolkata 700 064, India
\\[.5cm]
\end{center}
\vfill
\begin{abstract}
In this paper, triple neutral electroweak gauge boson production processes,
{\it viz.} $\gamma\gamma\gamma$, $\gamma\gamma Z$, $\gamma ZZ$ and $ZZZ$ 
productions merged to 1-jet have been studied at the leading order in QCD
in the context of Randall-Sundrum model at the LHC with center of mass energy 
$\sqrt{S}=13$ TeV. Decay of $Z$ bosons into lepton-pairs has been considered.
We present a selection of kinematical distributions matched 
to parton shower and show their deviation from the SM results
as a result of the RS model.  The uncertainties as a result of the
factorization and renormalization scales are also presented.
\end{abstract}

\vfill

\newpage
\section{Introduction}

The Standard Model (SM) of particle physics is now well established as the
true description of particles and their interactions at the experimentally
accessible energies.  The recently discovered 125 GeV scalar at the LHC
Run-I \cite{Aad:2012tfa,Chatrchyan:2012xdj}, behaves like the SM Higgs
 boson and this fixes the last free
parameter of the SM Lagrangian.  So far there is no indication of exotic
searches of beyond standard model (BSM) physics and the BSM scales have been
pushed further.  Remarkable agreement between the predicted SM values and
the measured cross sections spanning a broad range is a validation of the
analytical methods and the Monte Carlo tools developed to match the
challenges on the experimental and theoretical sides.  None the less the
SM is not a complete description of nature on various counts and the issue
could only be addressed from beyond \cite{Peskin:2015kka}.

Run-II at the LHC is now on with higher energies and luminosity.  Precision
measurement
of the properties of the newly discovered Higgs boson is a priority and
will be matched with precise higher order theoretical predictions to look
for any deviations from the SM.  Now that the Higgs boson is discovered
it is important to look at the massive vector boson scattering (VBS) cross
section which is uniterised by the Higgs boson.  The VBS gets contribution
from the non-abelian couplings of the electroweak gauge boson sector (a)
triple gauge boson coupling (TGC), (b) quartic gauge boson coupling (QGC)
and in addition, Higgs coupling to the massive gauge boson.  The TGC will
contribute to the di-boson final state and the Run-I has already placed
comparable limits to the anomalous couplings as the LEP experiments.  The
QGC coupling leads to tri-gauge boson final states, though the full process
would involve the TGC, fermion mediated processes and also Higgs mediated
processes.  The QGC has been reported for the first time by ATLAS 
collaboration using the VBS process, yielding a final state with 
two same sign W boson in association with two jets \cite{Aad:2014zda}
in a purely electroweak process and also measured the $W \gamma\gamma$ 
production cross section \cite{Aad:2015uqa} which is now accessible
with the 8 TeV LHC data set.  The CMS collaboration has also 
measured the QGC in the $WW\gamma$ and $WZ \gamma$ final states
\cite{Chatrchyan:2014bza}.  So far the observations
are consistent with SM predictions and as the sensitivity of these
measurements improves, 
the TGC and QGC that lead to tri-boson final states can not only test
the electroweak sector of the SM but also probe new physics.
In this paper we look at some of the tri-boson production processes merged
to 1-jet in the warped extra dimension model.

In the model proposed by Randall and Sundrum (RS) \cite{Randall:1999ee},
the non-factorizable geometry of the space-time with the inclusion of a
single warped extra spatial dimension, proposes a solution to the hierarchy
problem.  The SM fields are confined to a 3-brane, whereas
the gravity which propagates the full 5-dimensional space-time, manifests
as massive Kaluza-Klein (KK) modes in 4-dimensional space-time.  
RS model phenomenology of the virtual graviton
exchange have extensively been studied for gauge boson pair 
production processes {\it viz.}, $\gamma\gamma$ \cite{Kumar:2009nn},
$Z Z$ \cite{Agarwal:2009zg}, $W^+W^-$ \cite{Agarwal:2010sn} and also for
DY production \cite{Mathews:2005bw} at the next-to-leading order (NLO)
accuracy, because of their rich sensitivity to the model parameters.
This in turn helps to reduce the theoretical uncertainties whereby
constraining the RS model parameters. 
Recently the parton
shower effect has been considered for those processes in \cite{Das:2014tva}.
Nevertheless, the study of triple gauge boson production processes within
this model would also be phenomenologically important, as they could
effectively participate in interesting new physics searches at the TeV scale.
They have been studied in the SM at NLO 
\cite{Bozzi:2011en,Binoth:2008kt,Bozzi:2009ig,Lazopoulos:2007ix} level.
NLO results of the triple photon
production in the SM have recently been presented including the effect of
photon fragmentation \cite{Campbell:2014yka} or matching them with different
parton shower (PS) Monte Carlo programmes \cite{Mandal:2014vpa}.  These SM
processes also serve as potential backgrounds to a number of new physics
signals coming from different BSM scenarios. For example, the SM 
$\gamma\gamma\gamma$ process is a background to single photon production,
together with one techni-pion in technicolor model, whereas $\gamma\gamma
Z$ process in the SM is a background to the signal with di-photon plus
missing energy in gauge-mediated supersymmetric theories.

In this analysis,
we consider the production of neutral triple electroweak gauge bosons
in warped extra dimension model
at the LHC, {\it i.e.,} $PP\rightarrow VVV\ X$, where $V=\gamma,Z$ and $X$
denotes some hadronic final states. Similar processes have 
been analysed at LO in case of large extra dimensional models 
\cite{Kumar:2011jq}
\cite{Jiang:2012bf} \cite{Sun:2012bp}. In fact, study of these processes in
RS scenario bears equal importance, as their contributions in searching new
physics using the triple gauge boson productions are undeniable in
distinguishing physics arising from the potential BSM candidates 
like supersummetry or technicolor. 

This paper is organised in the following way: in section 2 we present
brief description of the RS model.
In section 3 we discuss the merging procedure and computational details.
Numerical results of various kinematical distributions
are provided in section 4 and finally we draw the conclusion of our study
in the last section. 

\section{Neutral Triple Vector Boson Production in RS Model}

In the RS model, the non-factorizable geometry is governed by the following
5-dimensional warped metric \cite{Davoudiasl:1999jd}, 
\begin{equation}
ds^{2} = e^{-2\kappa r_{c} |\phi|} \eta_{\mu\nu} dx^{\mu}dx^{\nu} - r_{c}^{2}d\phi^{2} \qquad ,
\end{equation}
where $\eta_{\mu\nu}$ represents the flat Minkowski metric. $\phi$ denotes
the fifth dimension ($0\leq\phi\leq\pi$) which is compactified on a $S^1/\mathbb{Z}_2$
orbifold with a radius $r_c$ and $\kappa$ is related to the curvature of the
$AdS_5$ space-time.  Two 3-branes with opposite tensions are situated on two
fixed points ($\phi=0,\pi$) of the extra dimension. The brane at $\phi=0$ is
called the \textquoteleft Planck brane\textquoteright\ and the other one at
$\phi=\pi$ is known as the \textquoteleft TeV brane\textquoteright. In the RS
scenario, all SM fields are considered to be confined on the TeV brane,
whereas the gravity can propagate in the full $4+1$ dimensions. The
interaction among the SM fields with the massive KK excitations
($h^{(n)}_{\mu\nu}$) of the graviton \cite{Giudice:1998ck, Han:1998sg} is
determined by the following Lagrangian, 
\begin{equation}
 \mathcal{L_{RS}} = - \frac{1}{\overline{M}_{Pl}} T^{\mu\nu} (x)h^{(0)}_{\mu\nu}(x) 
 - \frac{\overline{c}_{0}}{m_{0}} T^{\mu\nu} (x)\sum^{\infty}_{n=1} h^{(n)}_{\mu\nu}(x) \qquad ,
\end{equation}
where $\overline{c}_{0}=\frac{\kappa}{\overline{M}_{Pl}}$, $m_{0}=\kappa e^{-\kappa r_{c} \pi}$, 
$T^{\mu\nu}$ is the energy-momentum tensor for the SM particles and $\overline{M}_{Pl}$
is the reduced Planck scale.  Note that, the couplings of the zeroth KK mode to the SM
fields are $\overline{M}_{Pl}$ suppressed and hence this term can be practically neglected. 
However, the contribution from the higher modes with the coupling $\overline{c}_{0}/{m_{0}}$  
can be of the order of few TeV for a choice of $\kappa r_{c}\sim{\cal O}(10)$
\cite{Goldberger:1999uk, Goldberger:1999un} and they can produce significant observable
effects.  
The masses of the KK mode excitations are given by 
$M_n = x_n\,\kappa\,e^{-\pi\kappa r_c}$,
where $x_n$ indicates to the zeros of the Bessel function $J_1(x)$. 

The effective graviton propagator after summing over all the massive KK modes except the
zeroth one takes the following form \cite{Davoudiasl:2000wi} \cite{Kumar:2009nn}, 
\begin{eqnarray}
D_{eff}(s_{ij}) &=& \sum_{n=1}^{\infty} \frac{1}{s_{ij} - M_n^2 + i \Gamma_n M_n} \nonumber \\
&=& \frac{1}{m_{0}^{2}}\sum_{n=1}^{\infty} \frac{\left(x^{2}-x_{n}^{2}\right) 
-i x_{n}\frac{\Gamma_{n}}{m_{0}}}{\left(x^{2}-x_{n}^{2}\right)^{2}
+x_{n}^{2}\left(\frac{\Gamma_{n}}{m_{0}}\right)^{2}} \qquad ,
\end{eqnarray}
where $s_{ij}=(p_i+p_j)^2$, $x=\sqrt{s_{ij}}/m_0$ and $\Gamma_n$ denotes the
width of the resonance with mass $M_n$.  The total decay width of the graviton
can be calculated with the KK states decaying to the SM particles
 \cite{Han:1998sg} \cite{KumarRai:2003kk} in the following way, 
\begin{equation}
\Gamma_{n} = m_{0}\, \overline{c}_{0}^{2}\, x_{n}^{3}\, \Delta_{n} \qquad ,
\end{equation}
where $\Delta_{n}$ is given by, 
\begin{equation}
\Delta_{n} = \Delta_{n}^{\gamma\gamma} +\Delta_{n}^{ZZ} +\Delta_{n}^{WW} 
+\Delta_{n}^{HH} +\sum_{\nu}\Delta_{n}^{\nu\nu} 
+\sum_{l}\Delta_{n}^{ll} +\Delta_{n}^{gg}+\sum_{q}\Delta_{n}^{qq} \qquad .
\end{equation}
Here each $\Delta_{n}^{aa}$ corresponds to the coefficient coming from
the decay width calculation of the process $h^{(n)} \rightarrow a a$.
Unlike the large extra dimension model, the individual resonances of the 
graviton are well-separated in the RS model and they can be probed in 
invariant mass distribution.

The massive RS graviton could be produced in association with a photon
or a $Z$ boson and since the RS gravitons also couple to two photons or
two $Z$ bosons, the SM three neutral gauge boson final state distributions
could be altered due to the RS contributions and its interference with
the SM. 

\section{Merging Matrix Element with Parton Shower}

The leading order neutral triple gauge boson production processes
$PP \rightarrow VVV\ X$ at the LHC come from the subprocess,  
\begin{equation*}
q(p_{1}) +  \bar{q}(p_{2}) \rightarrow V(p_{3}) + V(p_{4}) + V(p_{5}) \qquad ,
\end{equation*}
where $V=\gamma, Z$ and $X$ is any final state hadron. This process has been 
merged with the 1-jet process $PP \rightarrow VVV j X$ in M{\sc ad}G{\sc raph}5
(MG) \cite{Alwall:2014hca}framework to
have a better description of different distributions. Due to extra radiation
emission, $q(\bar{q}) g$ initiated subprocesses also come up. The merged
events are then matched to a Parton Shower (PS).
The $Z$ bosons are let to decay to lepton pairs, thus accounting for the 
off-shell contributions.

In  LHC, additional jets are often produced from initial state radiation
and can alter the LO predictions for relevant observables.  Generally
these additional jets are simulated using PS monte carlo.  But these QCD
radiations in the PS programs are generated in the soft and collinear
approximation based on Sudakov form factors.  The widely separated and
hard emissions are not well-described in the PS approach, whereas the
fixed order tree level amplitudes can provide reliable predictions in
the hard region, but it fails in the collinear and soft limits.  Therefore
it is also essential to take into account the tree level amplitude
containing additional jets. Both descriptions have to be combined in an
appropriate matching method by avoiding double counting or gaps between
samples with different multiplicity.  Several algorithms have been proposed
for this purpose, mainly based on the event re-weighting (eg.\ CKKW)
\cite{Catani:2001cc}\cite{Krauss:2002up} or event rejection (eg. MLM)
\cite{Alwall:2007fs}.

The shower$\mbox{-}$k$_{T}$ scheme \cite{Alwall:2008qv}, based on event
rejection like MLM as implemented in M{\sc ad}G{\sc raph}5 is used in this
analysis.  In this scheme the events are generated by MG with a minimum
separation in the phase space $Q_{cut}$ and $P_{T_{min}}$ between the 
final$\mbox{-}$state partons ($ij$) and between the
final$\mbox{-}$state and initial$\mbox{-}$state partons ($iB$)
respectively which is characterized by the $k_{T}$ jet measure:
\begin{equation}
d_{ij}^{2} = min(p_{T_{i}}^{2},p_{T_{j}}^{2})\Delta R_{ij}^{2}   > Q_{cut}^2 ,
\hspace{1cm}    d_{iB}^{2} = p_{T_{i}}^{2} > p_{T_{min}}^{2} 
\end{equation}
Here
$\Delta R_{ij}^{2} = 2[\cosh(\eta_{i}-\eta_{j}) - \cos(\phi_{i}-\phi_{j})]$,
where $p_{T_i}$, $\eta_i$, $\phi_i$ are the transverse momentum,
pseudo-rapidity and azimuthal angle of the parton $i$.  The $k_T$
value is set as the renormalization scale at each QCD emission
vertex. The events are then passed to P{\sc ythia}\cite{Sjostrand:2006za} 
for showering. In shower-$k_T$ scheme, pythia $p_T$-ordered shower is
used for showering.  P{\sc ythia} reports the scale of the hardest
emission ($Q_{hardest}^{PS}$) in the shower and vetoes events based
on the $k_{T}$ values of the hardest shower emission instead of
performing a jet clustering and comparing to the ME.  If
$Q_{hardest}^{PS} > Q_{cut}$ for lower multiplicity samples,
then the event is rejected, whereas for highest multiplicity sample
an event is rejected if $Q_{hardest}^{PS} > Q^{ME}_{softest}$,
the scale of the softest parton in the event from ME.
We choose to work with $Q_{cut} = p_{T_{min}}$.

The fixed order merging approach gives a better description of the
region of hard and well separated  jet whereas the parton shower
takes care of the infrared region correctly.  These merged-matched
events provide a realistic framework to be compared to the
experimental outcomes.  The Lagrangian of the RS model is written
using FeynRules \cite{Alloul:2013bka} and it is combined together
with the SM Lagrangian.  The universal FeynRules output (UFO)
of the combined Lagrangian ({\it i.e.},
$\mathcal{L_{RS}}+\mathcal{L_{SM}}$) is then imported within
M{\sc ad}G{\sc raph}5 framework and used for the generation of
events.  The model parameters
$\overline{c}_0$ and $M_1$, the mass of the first excited KK
mode have been set as external inputs and we choose to work
with the following values: $M_1=1.7$ TeV and $\overline{c}_0=0.03$
which remain within the latest experimental bounds provided by
ATLAS \cite{Aad:2012cy} \cite{ATLAS:2011ab} and CMS
\cite{Khachatryan:2015sja}\cite{Chatrchyan:2011fq} collaborations.
In addition, we have systematically implemented the KK mode summation
algorithm in the spin-2 HELAS routine \cite{Hagiwara:2008jb}.
Kinematical distributions of various observables have been recalculated
for different di-final states such as, di-photon, Drell-Yan, $ZZ$, $W^+W^-$
in fixed order NLO  and NLO+PS using this
present layout and they are found to be in excellent agreement with 
those results that
are present in the literature
\cite{Kumar:2009nn,Agarwal:2009zg,Agarwal:2010sn,Mathews:2005bw,Das:2014tva}.
This essentially
ensures the proper execution of the whole computational set-up. We
have generated events for the following four neutral triple vector
boson production processes: ({\it i}) $\gamma\gamma\gamma$,
({\it ii}) $\gamma\gamma Z$, ({\it iii}) $\gamma ZZ$ and
({\it iv}) $ZZZ$ under the above mentioned arrangements. The $Z$ bosons
are decayed to lepton-pair which will be discussed in the next section
in detail.  Each of these processes consists of three types of contributions
coming from pure SM, pure RS and the interference between these two.


\section{Numerical Result}

In this section, we present numerical results of various kinematical
distributions for the above four processes.  All the results are
presented for LHC with center of mass energy $\sqrt{S} = 13$ TeV. In
our analysis, the following set of external parameters are used as input:
\begin{equation*}
m_{Z} = 91.188\ \mbox{GeV} \ ,
 \hspace{1cm}
\sin^{2}(\theta_{W}) = 0.222 \ ,
\end{equation*}
\begin{equation}
G_{F} = 1.16639\cdot 10^{-5}\ \mbox{GeV}^{-2} \ ,
\hspace{1cm}
\alpha^{-1} = 132.507
 \quad .
\end{equation}
During the generation of events we let the $Z$ bosons to decay to
$\ell^+ \ell^-$ pair.  Events are generated with loose cuts on the
transverse momentum $(P_T)$ and rapidity $(y)$ of the final state
particles:
\begin{equation}
\hspace{.5cm} P_{T}^{\gamma, \ell} > 15  GeV \ , \hspace{2cm} |y^{\gamma, \ell}|
\leq 2.6 \quad .
\end{equation}
The factorization scale ($\mu_{F}$) as well as the renormalization scale
($\mu_{R}$) are chosen as the invariant mass
of the final state particles and MSTW2008LO $(68\% CL)$ PDF has been used.
Throughout this paper, we have considered five massless quark flavors
($n_f$\,=\,$5$) and neglected all top quark contributions. In case of
processes containing two or three photons in final state, a photon
separation cut $\Delta R_{\gamma\gamma} > 0.3$ is used during event
generation.
$\Delta R_{\gamma\gamma} = \sqrt{(\Delta y)^{2}+ (\Delta \phi)^{2}}$
is the separation of the two photons in the rapidity-azimuthal angle
$(y,\phi)$ plane.  For processes involving leptons and photons in
 final state, we
have applied $\Delta R_{\gamma\ell} > 0.3$ and 
$\Delta R_{\ell^{+}\ell^{-}} > 0.3$.  For consistently merging 
0-jet sample with 1-jet sample we have followed the path
prescribed in \cite{Alwall:2008qv} and checked the stability of the
cross-section by varying the scale $Q_{cut}$.
Additional checks on the smoothness of the distributions, for example the
differential jet-rate (DJR) plots, in the ME and PS transition region have
been done.  For the four processes under consideration the choices of
$Q_{cut}$ are much the same around 90 - 95 GeV where we find best smooth 
DJR plots and also the matched cross-section remains within 13\% of 
the unmatched cross-section. For $\gamma \gamma \gamma$, $\gamma Z Z$
and $ZZZ$ we choose $Q_{cut} = 95$ GeV whereas for $\gamma \gamma Z$ we 
took $Q_{cut} = 90$ GeV.  Showering is done with P{\sc ythia}
$P_{T}$-ordered shower as described in the previous section.  Different
analysis cuts used on the final state particles at the time of
analysis are described in the following subsections for each processes.

For the processes involving photons, the photons can come from the hard
process or as a result of fragmentation which is a QED collinear effect.
In order to get rid of such collinear divergences without involving
additional non-perturbative effects, the smooth cone isolation criteria
on the photons as proposed by Frixione \cite{Frixione:1998jh}, is used.
A cone of radius 
$R = \sqrt{\left(y-y_{\gamma}\right)^{2}+\left(\phi-\phi_{\gamma}\right)^{2}}$
is considered in the ($y-\phi$) plane around a photon satisfying the condition
that the total hadronic transverse energy $E(R)$ within $R<R_{\gamma}$ would
be less than a maximum limit $E(R)_{max}$ given by,
\begin{equation}
E(R)_{max} = \epsilon_{\gamma}\, E_{T}^{\gamma}\, 
\left(\frac{1-\cos R}{1-\cos R_{\gamma}}\right)^{n} \qquad ,
\end{equation}
where $E_{T}^{\gamma}$ is the transverse energy of the photon; 
$\epsilon_{\gamma}, R_{\gamma}, n$ are three parameters of the Frixione
isolation. During event generation we choose 
$\epsilon_{\gamma} = 1, R_{\gamma} = 0.3, n =1$.

For the reconstruction of $Z$ bosons from opposite sign lepton-pair,
events are selected based on the selection criterion:
\begin{equation}
|M_{\ell^{+}\ell^{-}} -M_{Z}| \le  15 \quad GeV \qquad ,
\label{Z_id}
\end{equation}
where $M_{\ell^{+}\ell^{-}}$ is the reconstructed invariant mass of the
opposite sign lepton-pair. The following transverse momentum and 
pseudo-rapidity cut on jets during analysis in all four processes are
used
\begin{equation}
P_{T}^{j} > 50 \quad GeV,\hspace{1cm} \eta^{j} \le 4.5
\end{equation} 
For all four processes, an extra cut has been put on the
final state particles invariant mass ($M > 600$ GeV) for transverse
momentum distributions and rapidity distributions which are displayed
in the respective figures.  
We have checked that the generated events
give unbiased results with reference to the choices of
generation and analysis cuts.
Scale dependencies are obtained by varying the
renormalization scale ($\mu_{R}$) and the factorization scale ($\mu_{F}$)
\cite{Frederix:2011ss} in the range ($\mu_{R},\mu_{F}$) = 
($\kappa_{R} \mu_{0},\kappa_{F} \mu_{0}$), where $\mu_{0}$ is the invariant
mass of the three vector boson final states or its decay products when
$Z$s are involved.  The scale factors 
$\kappa_{R},\kappa_{F}$ are in the range $(1/2,2)$, we choose the combination 
($\kappa_{R}, \kappa_{F}) =$ (1/2,1/2), (1,1), (1/2,1), (1,1/2),
(1,2), (2,1), (2,2) to study the scale variation.
The scale uncertainty is represented by taking the envelope of all individual
variations.

\begin{figure}[!htb]
\centering{
\includegraphics[width=7.25cm,height=8.25cm,angle=0]{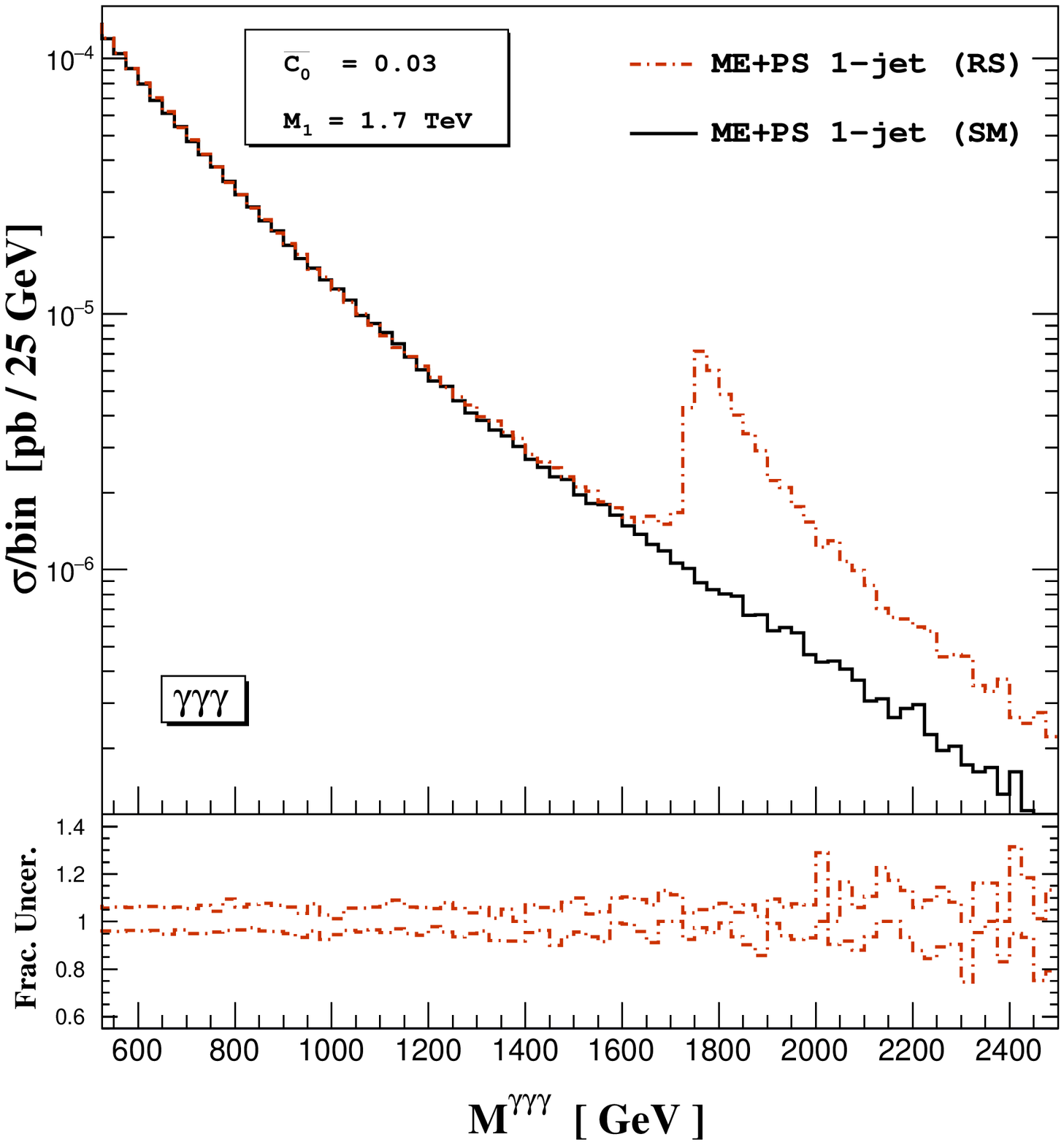}
\includegraphics[width=7.25cm,height=8.25cm,angle=0]{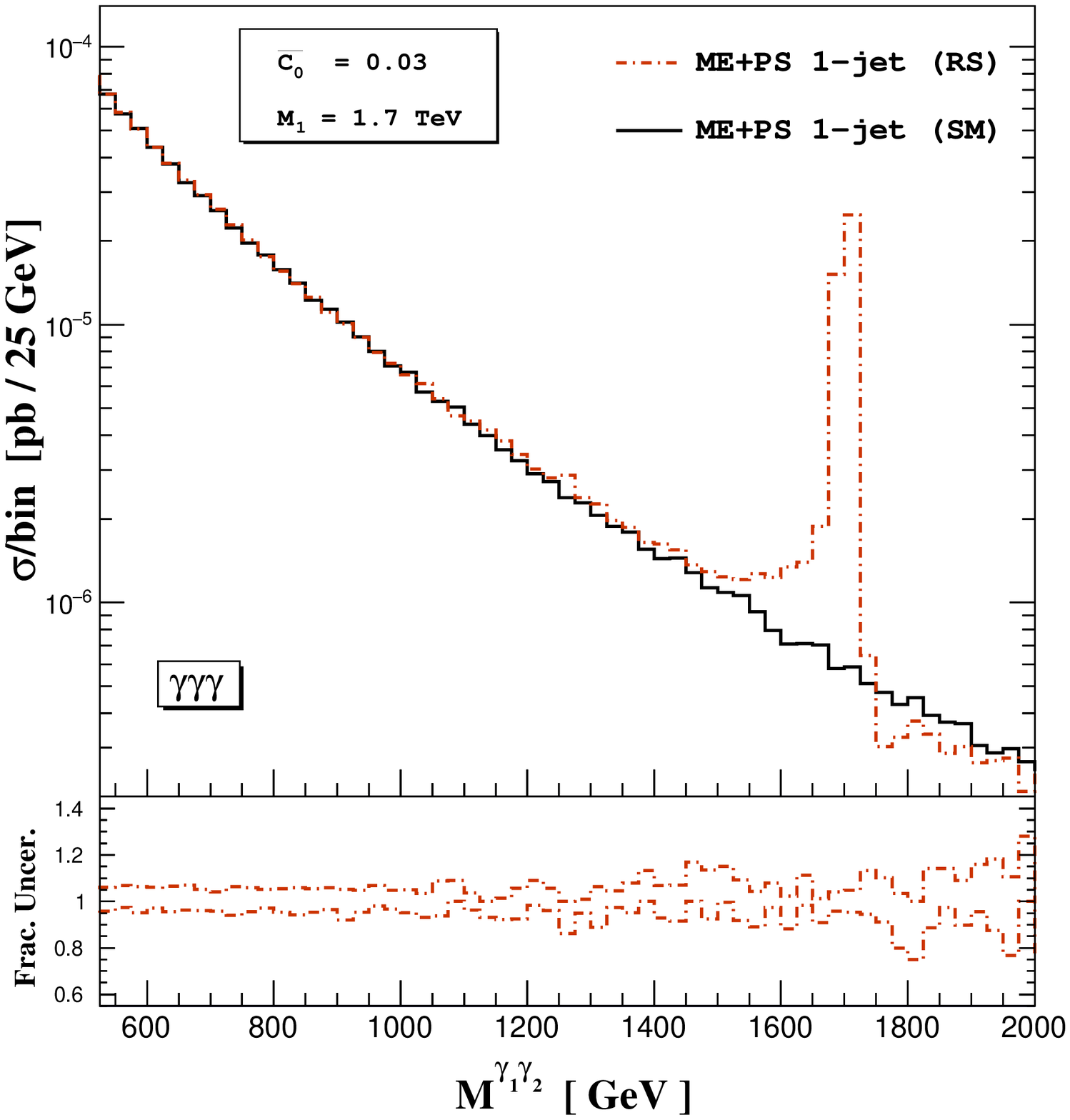}}
\caption{Invariant masses of $\gamma\gamma\gamma$ (left) and of two hardest
photons (right) for $\gamma \gamma \gamma$ production.}
\label{invmaaa}
\end{figure}

Next we discuss and present our results for each of the neutral triple boson
final states separately presenting some select distribution
that are of interest for the RS model.  For all figures, we followed
the following convention; we give the distributions corresponding to the
SM and SM+RS that contribute to the observable for ME+PS  merged to 1-jet
for central scale choice $(\kappa_R, \kappa_F)=(1,1)$.
In the lower inset we put the fractional scale uncertainty for ME+PS with
1-jet for RS case.

\subsection{$\gamma\gamma\gamma$}

Observing the $\gamma\gamma\gamma$ channel has a great advantage over the other 
triple neutral channels, because experimentally it provides a cleaner signature.
$\gamma\gamma\gamma$ production in the RS model has also been studied in
\cite{Atwood:2010gw}.
Here we present the result
merged with 1-jet as well as include showering thus improving the result, viable to
the experimental search.  During the analysis level we use more stringent cuts than
that used at the generation level:  (a) cuts on transverse momentum and 
pseudo-rapidity of final state photons used are $P_{T}^{\gamma} > 25$ GeV, 
$\eta^{\gamma} \le 2.5$,  (b) Frixione parameters used are
$R_{\gamma} = 0.4$, $\epsilon_{\gamma} = 1$, $n = 2$ and (c) photon-photon
separation cut $R_{\gamma\gamma} > 0.4$.  Finally the photons are ordered according
to their transverse momentum.
\begin{figure}[h]
\centering{
\includegraphics[width=7.25cm,height=8.25cm,angle=0]{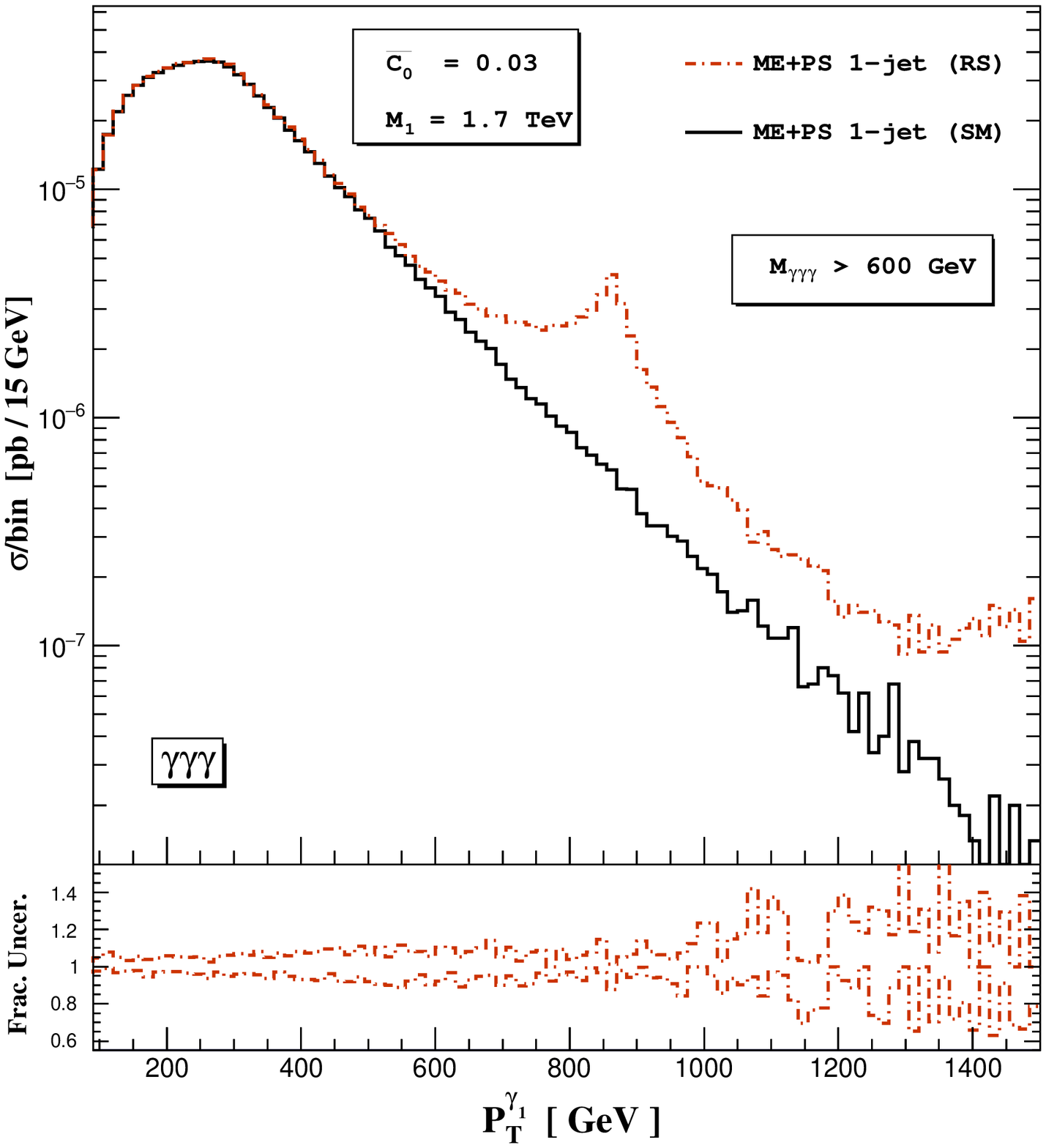}
\includegraphics[width=7.25cm,height=8.25cm,angle=0]{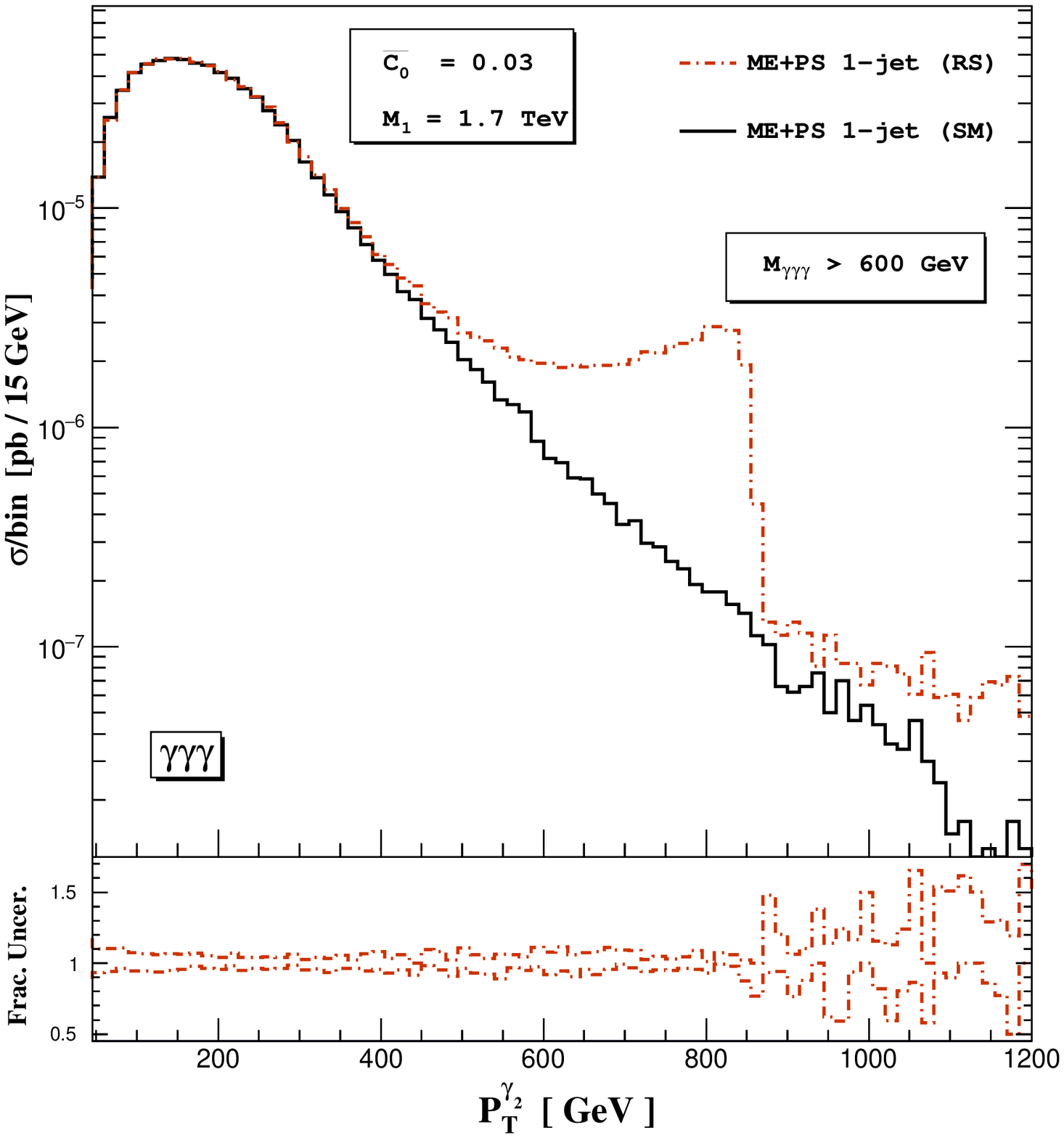}}
\caption{Transverse momentum distributions of hardest photon $\gamma_{1}$(left)
and next hard photon $\gamma_{2}$(right) for
$\gamma \gamma \gamma$ production.}
\label{ptaaa}
\end{figure}
\begin{figure}[!htb]
\centering{
\includegraphics[width=7.25cm,height=8.25cm,angle=0]{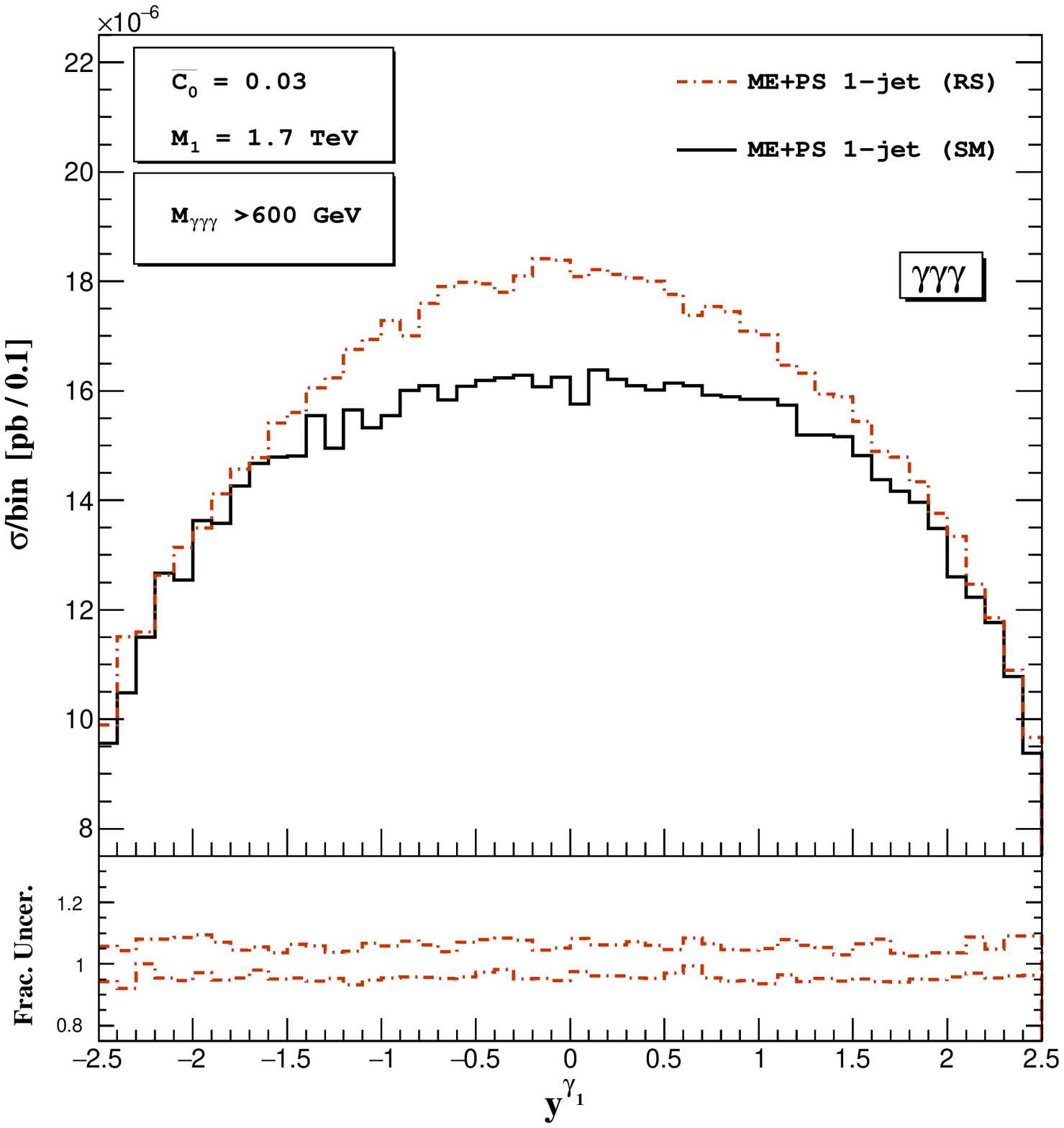}
\includegraphics[width=7.25cm,height=8.25cm,angle=0]{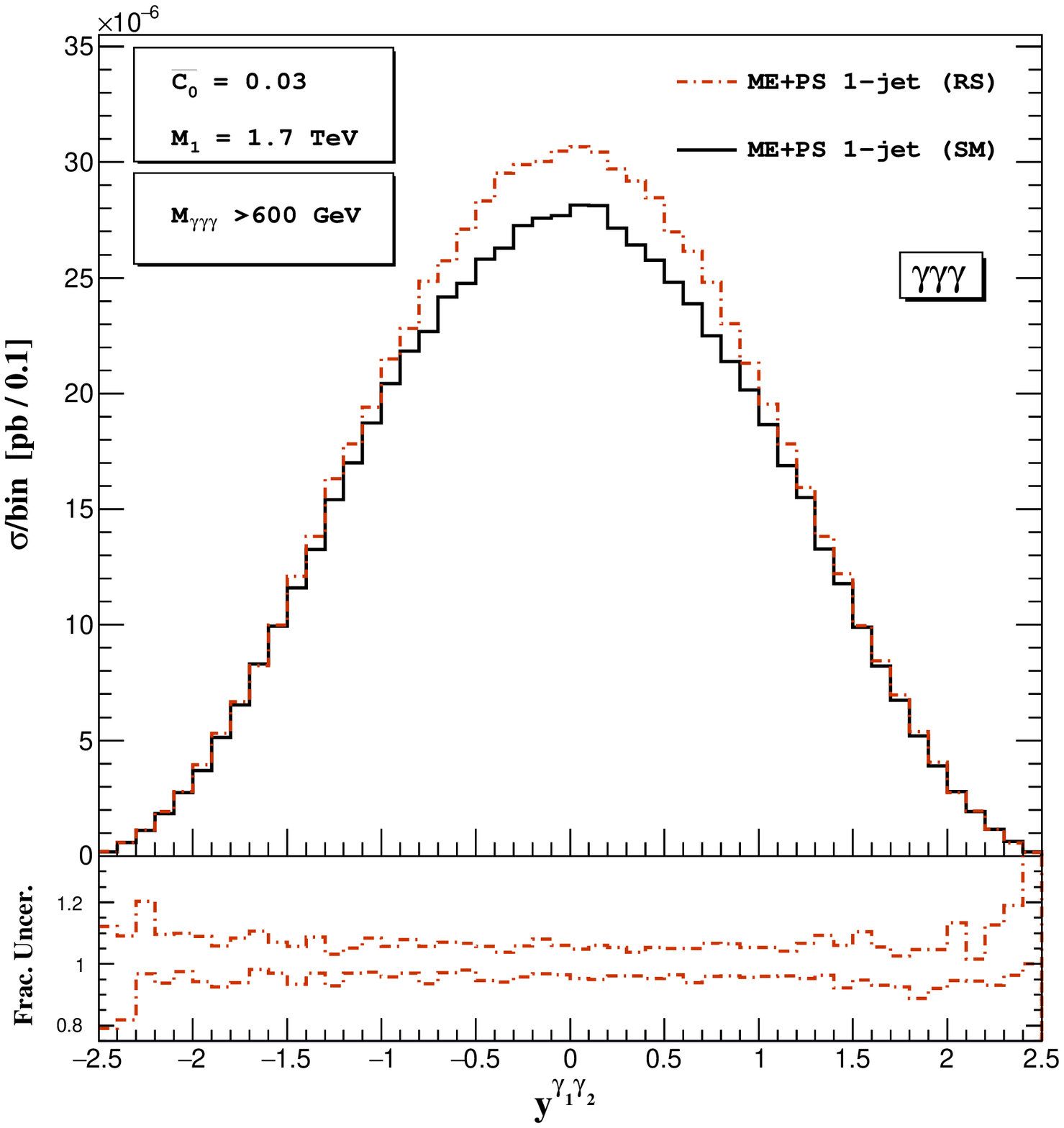}}
\caption{Rapidity distributions of  hardest photon (left) and hardest pair (right) for
$\gamma \gamma \gamma$ production.}
\label{rapaaa}
\end{figure}

In the Fig.~\ref{invmaaa}, we present the invariant mass distribution for 
tri-photon on the left panel and the invariant mass of the hardest
di-photon state on the right panel.  In the invariant mass distribution
of the tri-photon the peak appears near 1.7 TeV. The peak is slightly shifted
from 1.7 TeV towards higher invariant mass region. This is evident from the
fact that the RS graviton is produced in association with a vector boson.
In the expression for invariant mass there
are dot products between all three momenta of the final state photons,
thus shifting the peak slightly towards the higher invariant mass region.
A cleaner signature of RS graviton can be found in the invariant mass
distribution of the hardest two photon pairs where the RS peak appears at
1.7 TeV. From the transverse momentum distributions (Fig.~\ref{ptaaa})
as well as in the rapidity distributions (Fig.~\ref{rapaaa}),
significant deviation from SM results is observed.
In general the 1-jet merged sample gives a harder distributions.
For the Fig.~\ref{invmaaa} (left panel) the uncertainty at the RS
peak is about 7.7\% and for the
di-photon invariant mass (right panel) the uncertainty is about 10\%.
For the $p_T$ distribution the uncertainties are larger, for the
hardest photon (Fig.~\ref{ptaaa}) it is about 15\% and for the second hardest 
it is about 8.3\% around the peaks in the $P_T$ distribution which 
correspond to about half the RS resonance.  For the rapidity plots (Fig.~\ref{rapaaa})
the uncertainty is about 10\% in the central rapidity region.

From the invariant mass distributions (Fig.~\ref{invmaaa}) an
estimation of the signal and background events for the LHC Run-II can be made.
For this purpose we have considered the invariant mass distributions
over 1500 GeV where the enhancement due to RS graviton signal is 
clearly visible over the SM background. We find that for $100~fb^{-1}$
luminosity there are 8 signal events over 3 background events from the
$M^{\gamma\gamma\gamma}$ distribution (Fig.~\ref{invmaaa} left panel)
and 6 signal events over 2 background events from $M^{\gamma_{1}\gamma_{2}}$ 
distribution (Fig.~\ref{invmaaa} right panel). 
\subsection{$ \gamma\gamma Z $}

\begin{figure}[h]
\centering{
\includegraphics[width=7.25cm,height=8.25cm,angle=0]{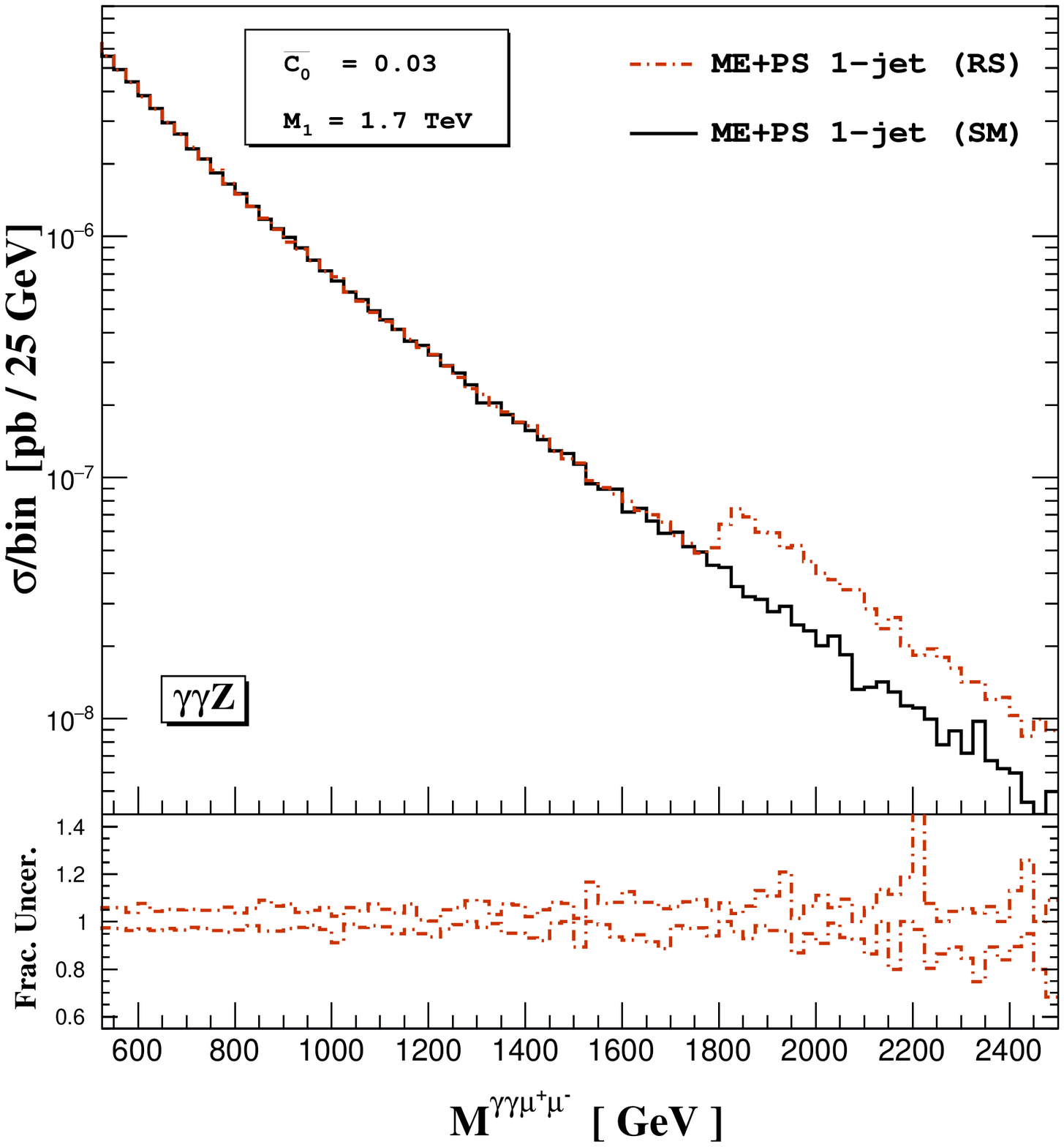}
\includegraphics[width=7.25cm,height=8.25cm,angle=0]{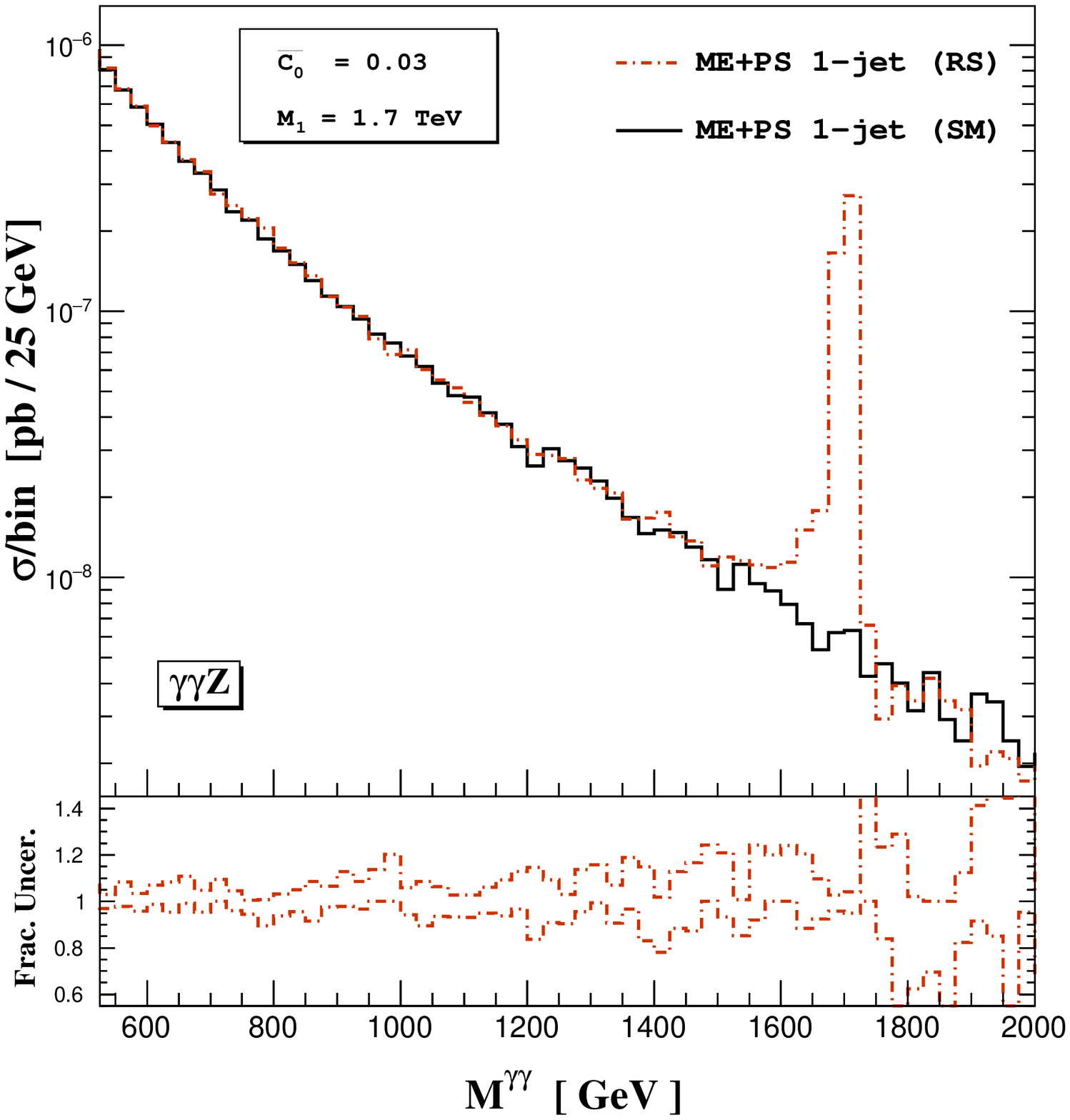}}
\caption{Invariant mass distributions of $\gamma\gamma \mu^{+}\mu^{-}$ (left)
and $\gamma\gamma$(right) for
$\gamma \gamma Z$ production.}
\label{invmaaz}
\end{figure}
\begin{figure}[!htb]
\centering{
\includegraphics[width=7.25cm,height=8.25cm,angle=0]{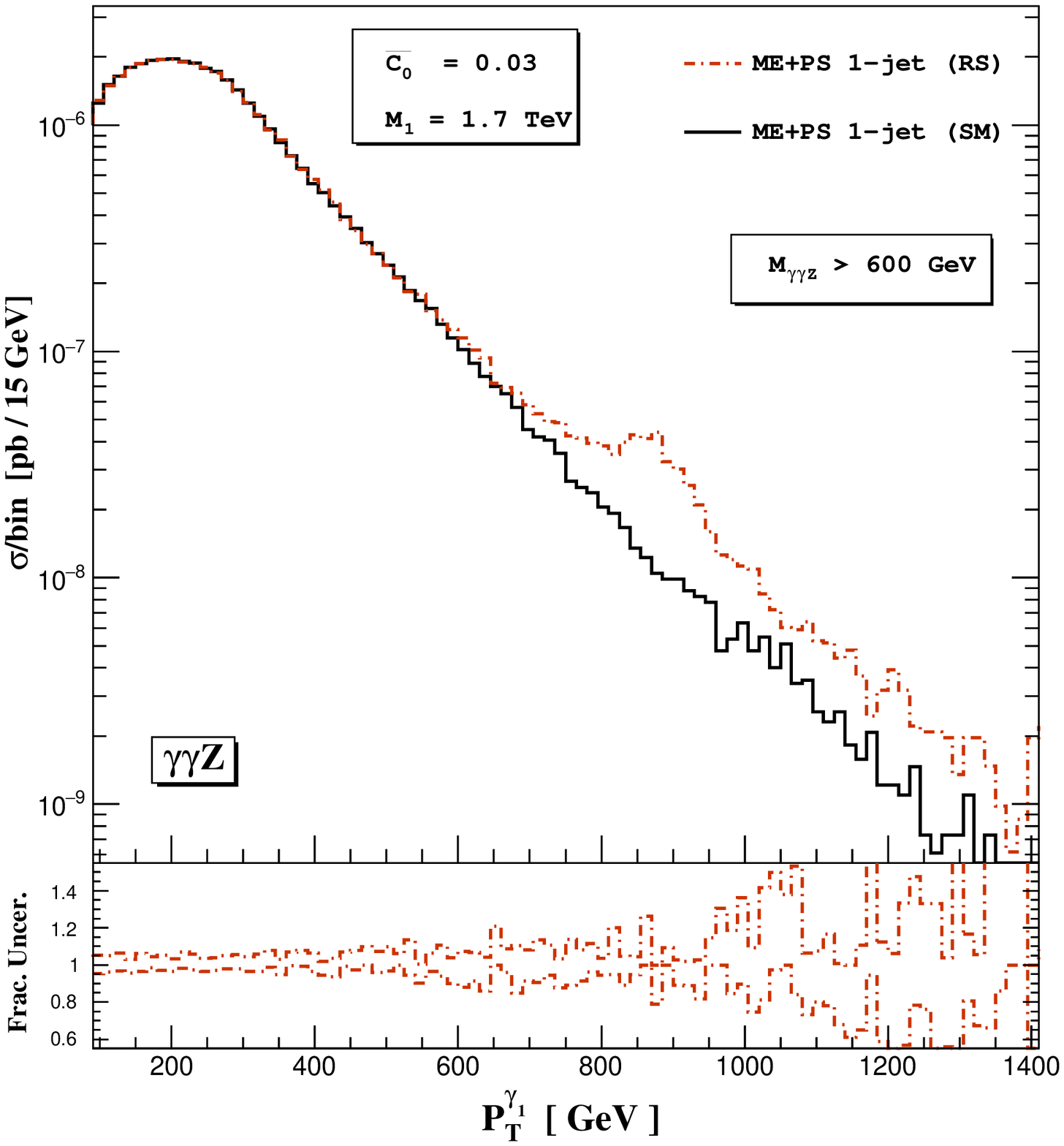}
\includegraphics[width=7.25cm,height=8.25cm,angle=0]{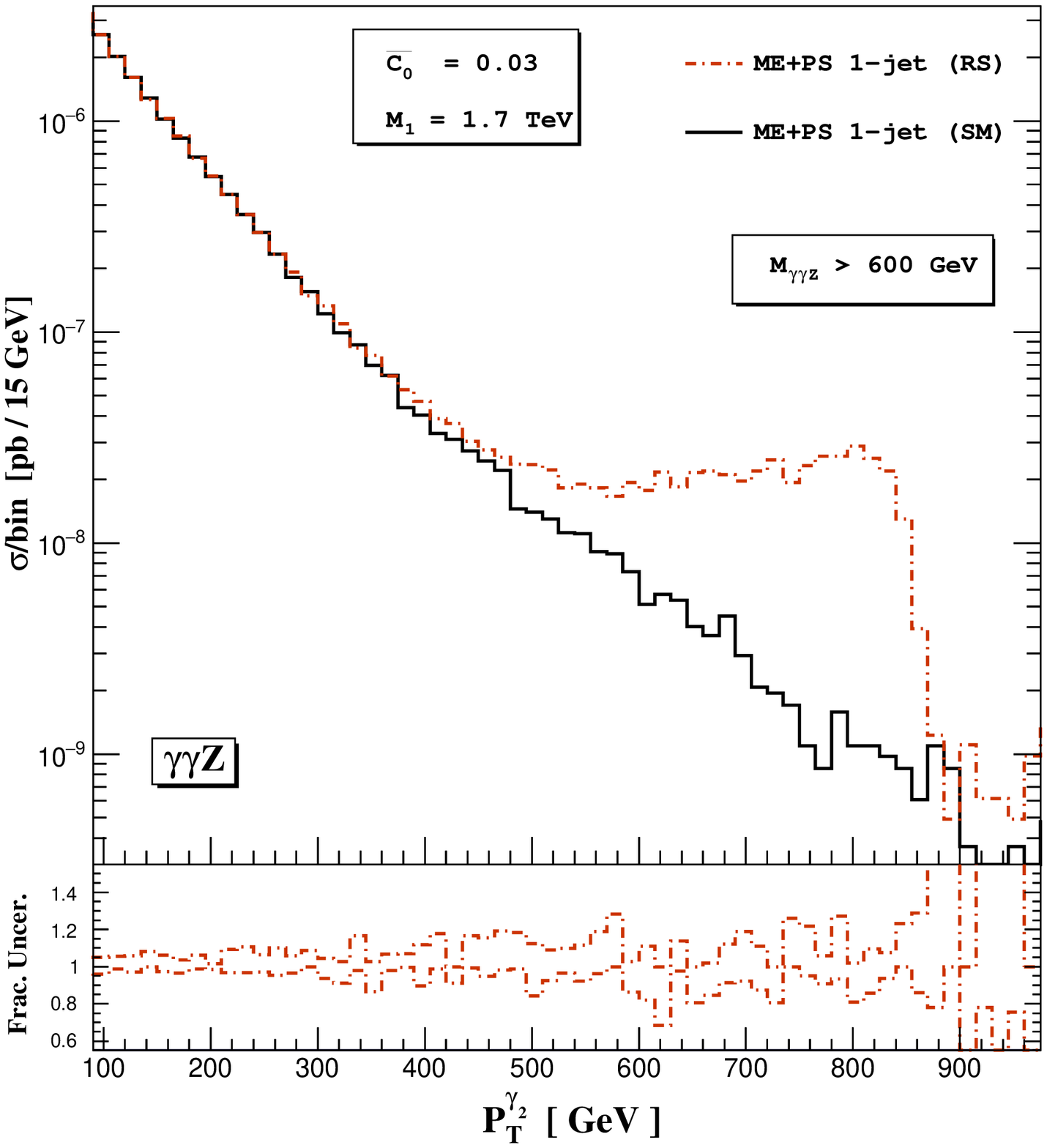}}
\caption{Transverse momentum distributions of hardest photon $\gamma_{1}$(left)
and second hard photon $\gamma_{1}$(right) for
$\gamma \gamma Z$ production.}
\label{ptaaz}
\end{figure}

In $\gamma \gamma Z$ production the effect of the massive RS KK-modes can be
observed in the $\gamma \gamma Z$ invariant mass as well as in the di-photon
invariant mass (Fig.~\ref{invmaaz}).  The $Z$ boson is allowed to decay to $\mu^+ \mu^-$ 
pair.  Minimal analysis level cuts on the final state leptons and photons
transverse momentum 
$P_{T}^{\gamma, l} \geq 25$ GeV and  pseudo-rapidity cut $\eta^{\gamma,l} \le 2.5$
are put.  In order to fulfill detector resolution, cuts on the
separation of photon, lepton and jets are imposed:
\begin{equation}
\hspace{1cm} R_{\gamma l} > 0.4, \hspace{1cm} R_{ll} > 0.4, 
\hspace{1cm} R_{l j} > 0.7 \quad .
\end{equation}
The invariant mass distribution of $\gamma\gamma \mu^{+}\mu^{-}$ system
shows an enhancement at about the RS peak but is additionally shifted 
compared to the tri-photon case due to the mass of the $Z$ boson.
In this process the RS graviton can only decay into a photon-pair,
hence the invariant mass distribution of di-photon pair shows the peak at the RS mass. 
The uncertainty in the enhanced region is about 8\% for the 
$\gamma\gamma \mu^{+}\mu^{-}$ invariant mass distribution and for the
di-photon invariant mass at the RS peak it is about 9.9\%.
The transverse momentum distributions (Fig.~\ref{ptaaz}) of each photon
clearly shows the RS signature appearing nearly at the half of the KK mass.
The uncertainty for the transverse momentum distribution are much higher,
at about 26\%.

In the $P_T$ distributions in Figs.\ \ref{ptaaa}, \ref{ptaaz}
the parton shower resums the large logarithms in the collinear
region which 
suppress the low $P_T$ cross section.  In this region the scale
uncertainty is also low as the higher logarithms are resummed
to all orders.

\subsection{$ \gamma Z Z $}

\begin{figure}[!htb]
\centering{
\includegraphics[width=7.25cm,height=8.25cm,angle=0]{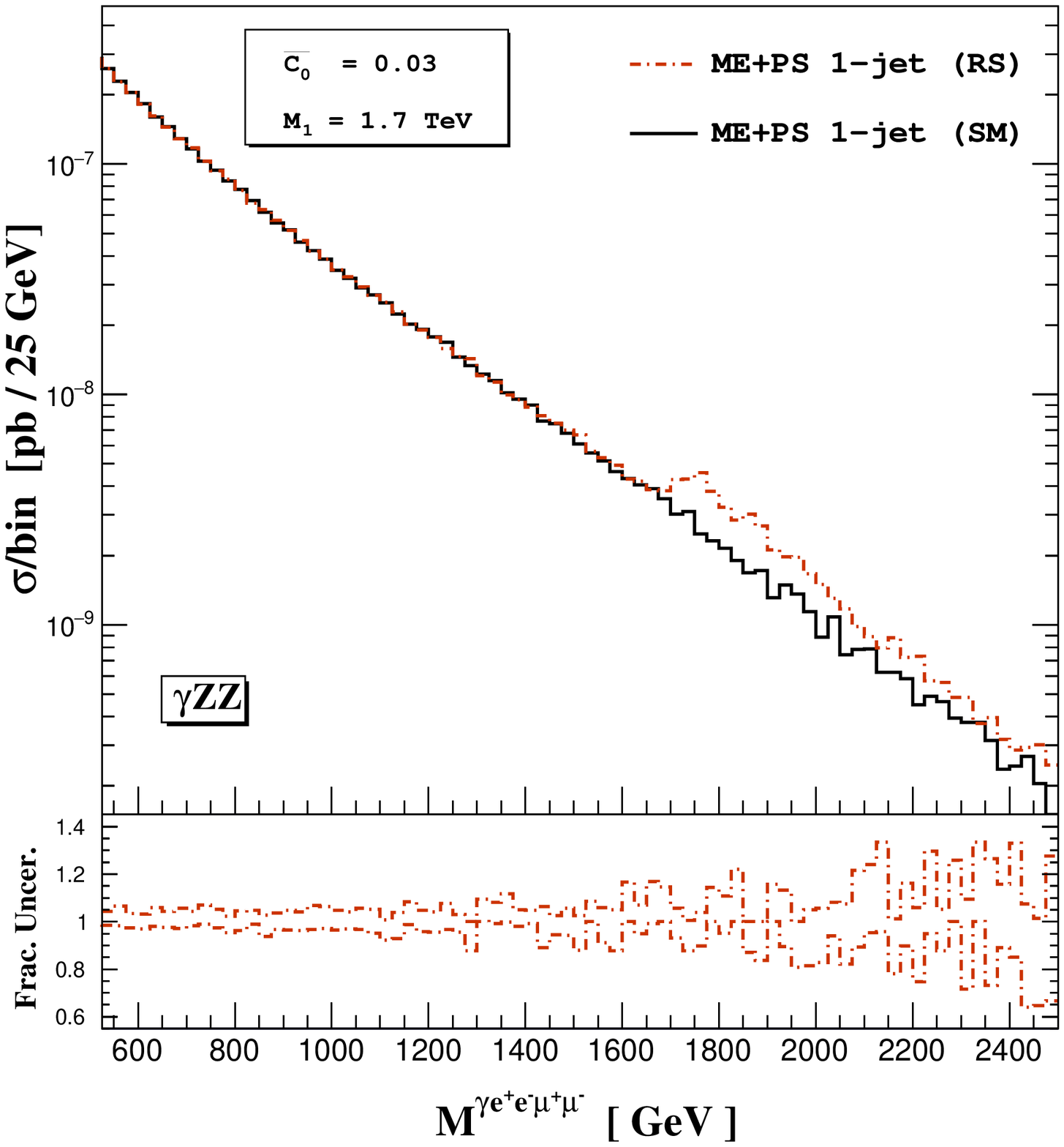}
\includegraphics[width=7.25cm,height=8.25cm,angle=0]{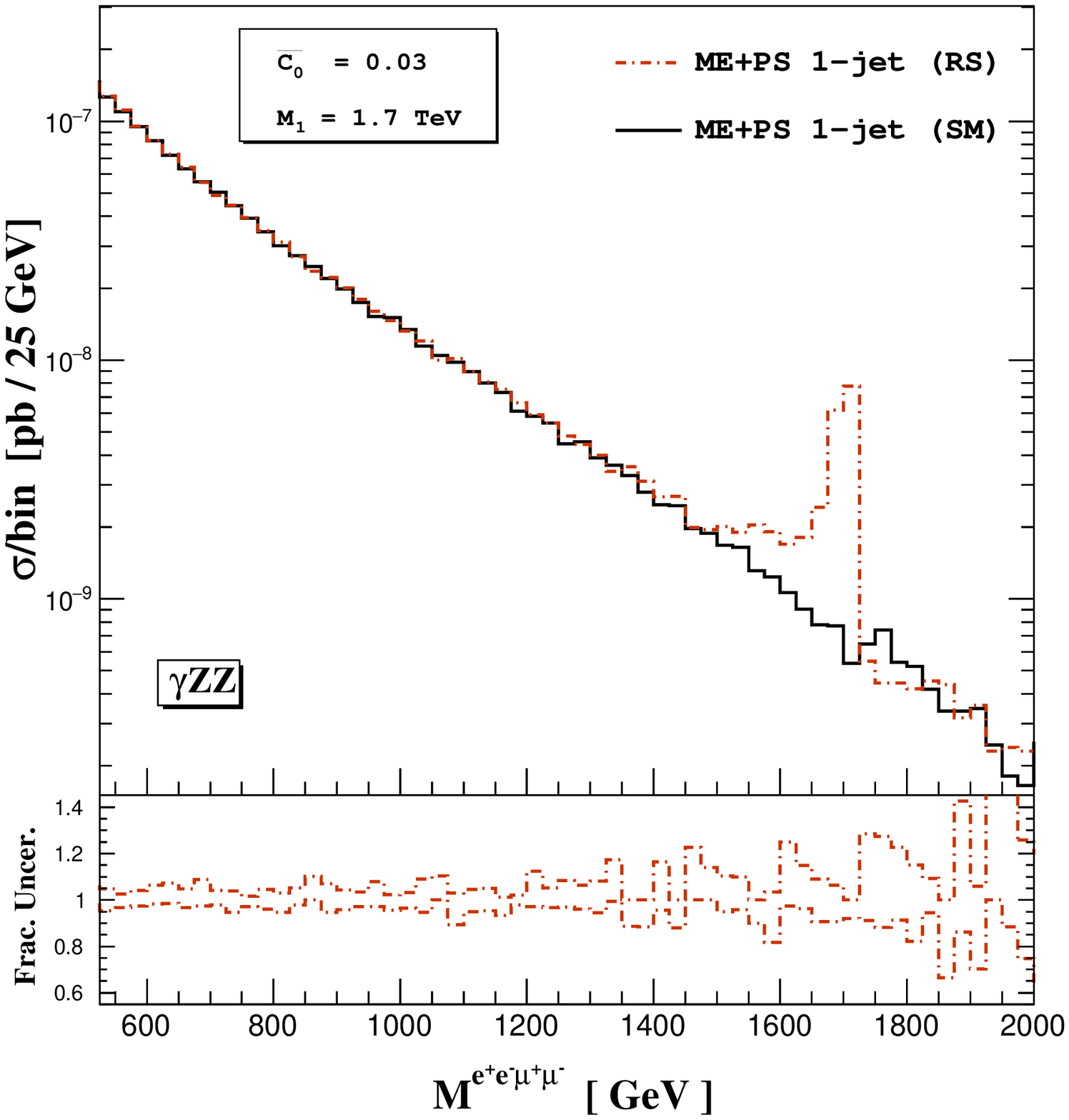}}
\caption{Invariant mass distributions of $\gamma e^{+}e^{-}\mu^{+}\mu^{-}$ (left) 
and  $e^{+}e^{-}\mu^{+}\mu^{-}$  (right) for 
$\gamma Z Z$ production.}
\label{invmazz}
\end{figure}

The same set of analysis cuts are used here, as the $\gamma\gamma Z $ case.
One $Z$-boson is decayed to a $e^{+}e^{-}$ pair and the other $Z$-boson to
a $\mu^+ \mu^-$ pair during event generation.  In the $\gamma Z Z$ invariant
mass distribution (Fig.~\ref{invmazz}) there is slight enhancement of RS 
contribution over the SM, around the invariant mass region (1.7 TeV).  But the
effect of massive KK state is best found on the di-$Z$ boson invariant mass
distribution (Fig.~\ref{invmazz}) and it peaks at 1.7 TeV as expected. 
The $p_T$ distributions are not very useful to discriminating the 
RS signatures.  The uncertainty for the invariant mass distribution 
of $\gamma e^{+}e^{-}\mu^{+}\mu^{-}$ system is about 15\% and for the invariant
mass of the $e^{+}e^{-}\mu^{+}\mu^{-}$ system which is as a result of
the RS graviton decay is about 8.8\%. 

\subsection{$ZZZ$}

\begin{figure}[!htb]
\centering{
\includegraphics[width=7.25cm,height=8.25cm,angle=0]{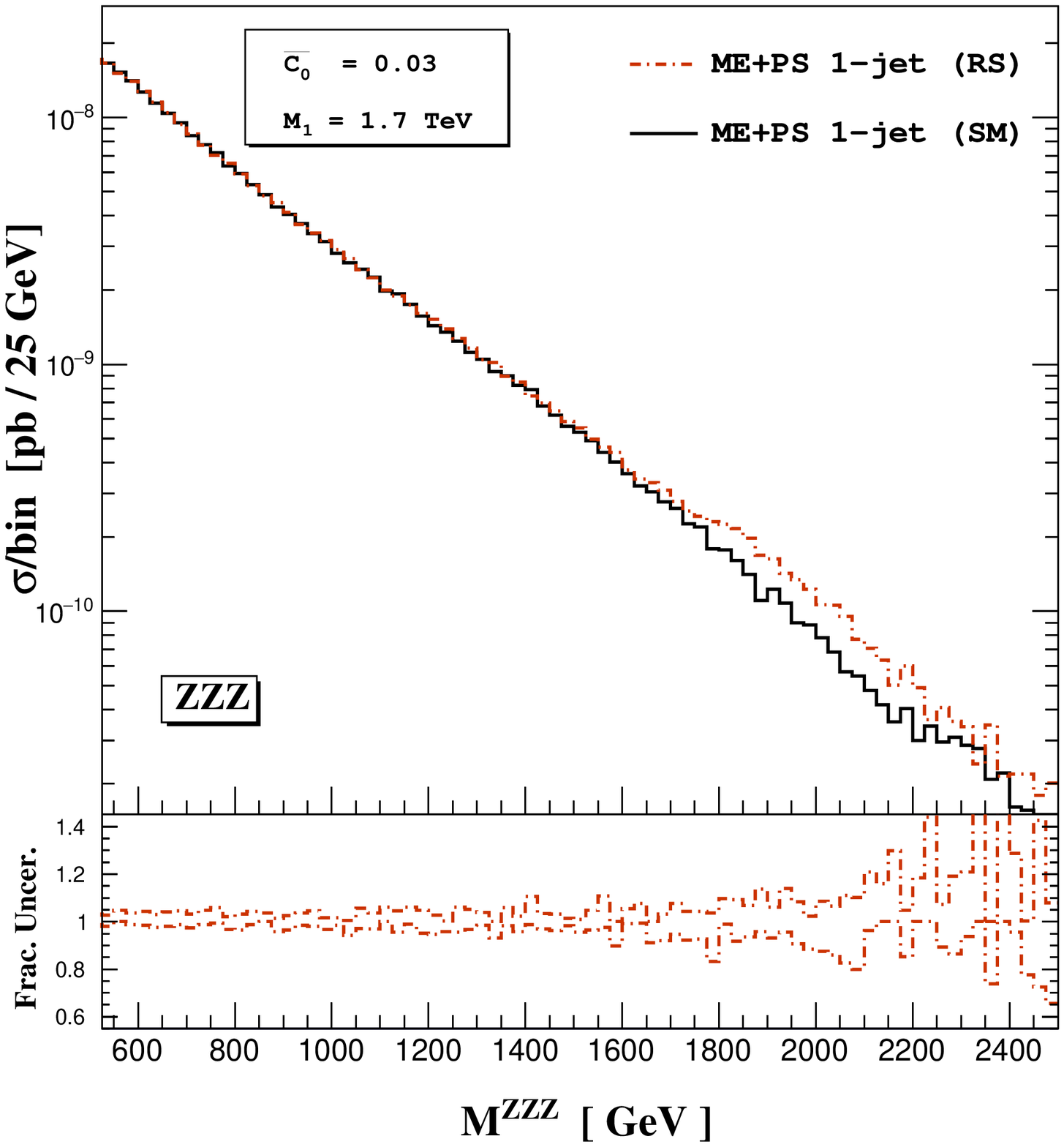}
\includegraphics[width=7.25cm,height=8.25cm,angle=0]{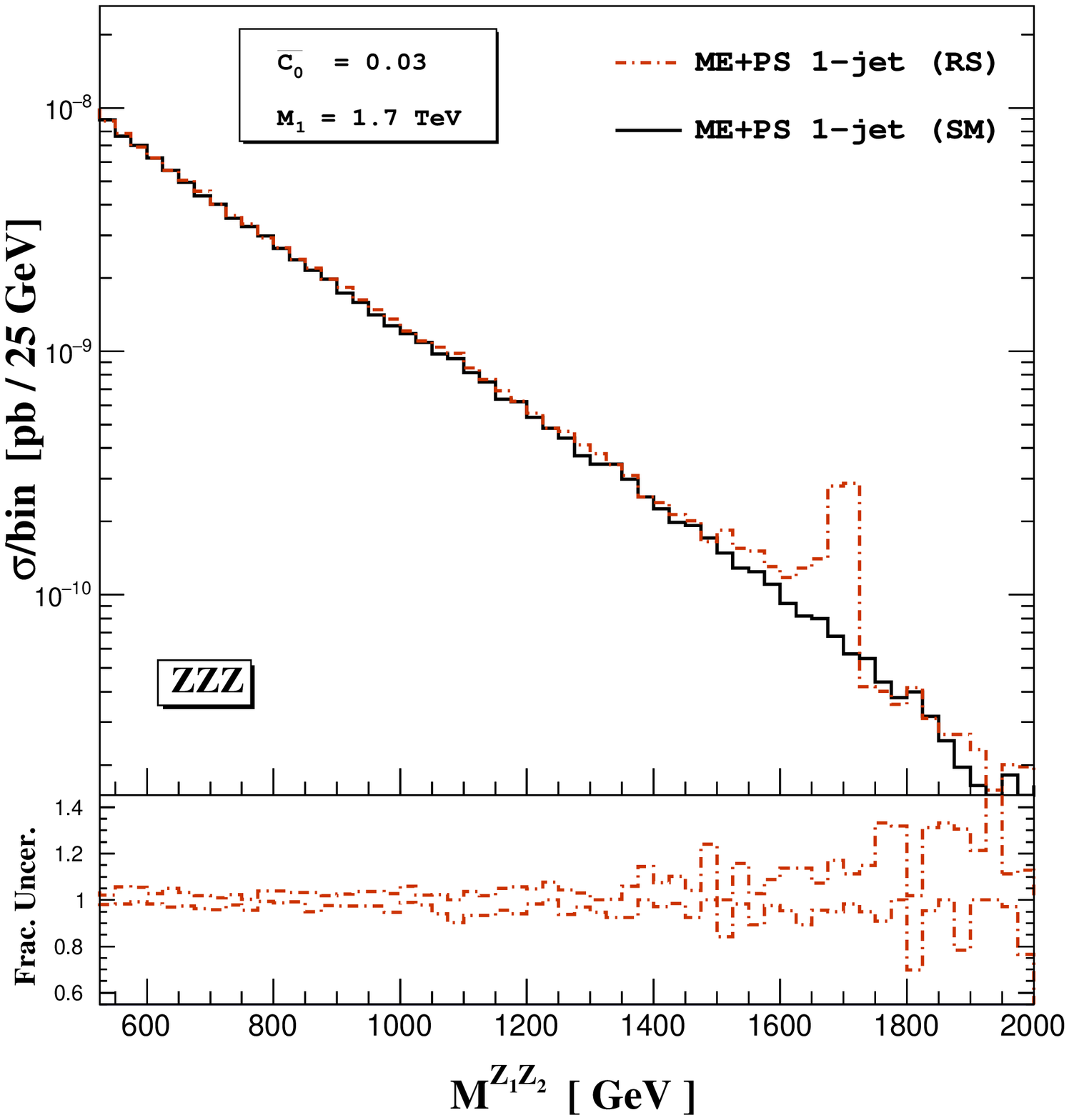}}
\caption{Invariant mass distributions of $e^{+}e^{-}\mu^{+}\mu^{-}\mu^{+}\mu^{-}$ 
(left) and hardest two lepton pairs (right) for $ZZZ$ production.}
\label{invmzzz}
\end{figure}

For $ZZZ$ production, we choose to decay two $Z$ bosons to two
$\mu^{+}\mu^{-}$ pairs whereas the other $Z$ boson is decayed to a
$e^{+}e^{-}$ pair.  For this process also same set of 
analysis cuts has been used, $P_{T}^{l} \geq 25$ GeV, $\eta^{l} \le 2.5$, 
where $l = e^+,e^-,\mu^+,\mu^-$.
The $Z$ bosons are reconstructed according to the criterion described at the
beginning of this section (see Eq.\ (\ref{Z_id})).  The reconstructed $Z$
bosons are then ordered according to their transverse momentum.
Thus in the figures $Z_{1,2}$
represents transverse momentum ordered $Z$ bosons. Small enhancement
over the SM can be seen in the $ZZZ$ invariant mass distribution 
(Fig.~\ref{invmzzz} left panel).
The uncertainty in that region is found to be about 13-14\%.
In this case also the RS contribution is best found in the hardest
two $Z$ boson invariant mass distribution (Fig.~\ref{invmzzz} right panel). 
The uncertainty at the peak region is about 13.25\%.

\section{Conclusion} 

In the context of RS model, effects of the exchange of virtual KK graviton
have been studied on the neutral triple gauge boson production processes
at the 13 TeV LHC.  This process could play
a vital role in discriminating physics beyond the SM and in estimating
the contribution coming from potential BSM scenarios in new physics
searches.  We have incorporated the RS model, using FeynRules in
association with an algorithm that takes care of the KK mode summation
of gravitons within M{\sc ad}G{\sc raph}5 environment and performed a
number of checks to ensure their proper implementation. 

We have merged $ P ~P \to VVV$ and $P ~P \to VVV +j$
event samples for better prediction of the
distributions and observe that it gives harder distributions compared to
the unmerged sample. To make theoretical prediction closer to the
experimental situation, we have also matched the merged events with parton
shower. Final state $Z$ bosons are allowed to decay to either of the
following leptonic decay modes: ({\it i}) $Z\rightarrow e^+e^-$, ({\it ii})
$Z\rightarrow \mu^+\mu^-$ at the time of event generation, thus taking into
account off-shell effects as well.  For process with more than one stable
photon, the photons are ordered according to their $P_T$ and then the
required number of photons are collected, based on their hardness. Likewise,
for triple $Z$-boson production, $Z$ bosons are reconstructed from their
daughter particles and then ordered according to the hardness. Numerical
results of some selective differential distributions for a set of kinematical
variables have been presented for the merged samples.
All these codes are flexible enough to incorporate the experimental cuts,
different values for model parameters {\it etc.} and they can be used to
obtain numerical results of any kind of distributions that would indeed
help the experimental collaborations. 

Of the neutral tri-gauge boson final states considered here, the tri-photon 
final state has the highest rate and can be used to look for 
signatures of the RS model.  For the tri-photon process, the invariant
mass, $P_T$ distributions of various photons ordered in terms of the
hardness and rapidity distributions are all good discriminators of the
RS model.  
The scale uncertainties 
are by and large within 10\% for the invariant mass distribution.
In the tri-final invariant mass distribution the cross section 
is enhanced in the RS resonance peak region, which diminishes in going from
the $\gamma\gamma\gamma$ to the $ZZZ$ final state.
The di-invariant mass distribution for all four processes
is a clear indicator of the RS resonance peak. In the 
di-invariant mass distributions the peak of the RS resonance is 
most enhanced for the  $\gamma\gamma\gamma$ and
diminishes in going to the $ZZZ$ production
process.

\section*{Acknowledgements}
We acknowledge the high performance cluster computing facility of
the Theory Division, SINP. The authors also thank Satyajit Seth
for collaboration during the initial stages of this work.  GD
would like to thank Olivier Mattelaer, Paolo Torrielli for discussion
regarding M{\sc ad}G{\sc raph}5.  GD also thanks Suvankar Roy Chowdhury,
Atanu Modak of CMS for useful discussions.  The authors would like to
thank V.\ Ravindran for useful comments.  Work of GD is supported by 
funding from Department of Atomic Energy, India.

\bibliographystyle{JHEP}
\bibliography{VVVRS_v4}
\end{document}